\RequirePackage{fixltx2e}
\documentclass[twocolumn,floatfix,showpacs,preprintnumbers,nofootinbib%,superscriptaddress
]{revtex4-1}
\usepackage[english]{babel}
\usepackage[utf8x]{inputenc}
\usepackage{mathrsfs}
\usepackage{amssymb}
\usepackage{amsmath}
\usepackage{mathtools}
\usepackage{mathrsfs}
\usepackage{graphicx}
\usepackage{fixltx2e}

\usepackage{xcolor}
\definecolor{lcolor}{rgb}{0.5,0,0}
\definecolor{citcolor}{rgb}{0,0.3,0.0}

\usepackage[breaklinks,colorlinks,urlcolor=blue,citecolor=citcolor,linkcolor=lcolor]{hyperref}
\usepackage{subdepth}

\newcommand{\rt}{{\mathbf{r}}}
\newcommand{\xt}{{\mathbf{x}}}
\newcommand{\bt}{{\mathbf{b}}}

\newcommand{\yt}{{\mathbf{y}}}
\newcommand{\zt}{{\mathbf{z}}}
\newcommand{\pt}{{\mathbf{p}}}
\newcommand{\qt}{{\mathbf{q}}}
\newcommand{\kt}{{\mathbf{k}}}

\newcommand{\ptt}{p_\perp} % scalar
\newcommand{\ktt}{k_\perp} % scalar
\newcommand{\qtt}{q_\perp} % scalar
\newcommand{\ltt}{l_\perp} % scalar
\newcommand{\lt}{\mathbf{l}}

\newcommand{\xif}{{\xi}_\text{f}}
\newcommand{\xf}{{x}_\text{f}}

\newcommand{\ud}{\mathrm{d}}

\newcommand{\tr}{\, \mathrm{Tr} \, }

\newcommand{\nc}{{N_\mathrm{c}}}

\newcommand{\half}{\frac{1}{2}}

\newcommand{\cf}{C_\mathrm{F}}

\newcommand{\nr}[1]{(\ref{#1})}

\newcommand{\stt}{S_\perp}

\newcommand{\qs}{Q_\mathrm{s}}

\newcommand{\as}{\alpha_{\mathrm{s}}}

\newcommand{\eq}{Eq.~}

\newcommand{\fcal}{\mathcal{F}}
\newcommand{\ical}{\mathcal{I}}
\newcommand{\jcal}{\mathcal{J}}
\newcommand{\scal}{\mathcal{S}}
\newcommand{\kcal}{\mathcal{K}}

\allowdisplaybreaks

\begin{document}

\author{B. Duclou\'e}
\author{T. Lappi}
\author{Y. Zhu}
\affiliation{
Department of Physics, P.O. Box 35, 40014 University of Jyv\"askyl\"a, Finland
}
\affiliation{
Helsinki Institute of Physics, P.O. Box 64, 00014 University of Helsinki,
Finland
}

\title{
Single inclusive forward hadron production at next-to-leading order
}


\preprint{}

\begin{abstract}
We discuss single inclusive hadron production from a high energy quark scattering off a strong target color field in the Color Glass Condensate formalism. Recent calculations of this process at the next-to-leading order accuracy have led to negative cross sections at large transverse momenta. We identify the origin of this problem as an oversubtraction of the rapidity divergence into the Balitsky-Kovchegov evolution equation for the target. We propose a new way to implement the kinematical restriction on the emitted gluons to overcome this difficulty.
\end{abstract}

\maketitle

\section{Introduction}
Hadronic reactions at modern collider energies reach a kinematical domain where gluon densities can be nonperturbatively large even at short distance scales where the QCD coupling is weak. A convenient effective theory of QCD in this regime is provided by the Color Glass Condensate (see e.g. \cite{Gelis:2010nm}), which describes the nonlinear small $x$ degrees of freedom in a hadron or nucleus as a classical color field. An ideal way to study these dense color fields is to probe them with simple dilute probes in a high energy collision, such as in deep inelastic scattering or forward particle production in proton-nucleus collisions. In the latter case the dilute probe is provided by the relatively well understood large $x$ partons of the probe proton which, at forward rapidity, scatter off the small $x$ color field of the target.

Several calculations~\cite{Dumitru:2005gt,Albacete:2010bs,Tribedy:2011aa,Rezaeian:2012ye,Lappi:2013zma} of forward single inclusive particle production in this framework have provided a good description of available experimental data using the leading order expression for the cross section~\cite{Dumitru:2002qt}. As is often the case in QCD, the  leading order calculations leave the overall normalization of the cross section quite uncertain. It would, therefore, be desirable to systematically go to higher orders in perturbation theory in these cross section calculations. Similar developments towards higher order have recently taken place concerning the (Balitsky-Kovchegov (BK)~\cite{Balitsky:1995ub,Kovchegov:1999yj} or JIMWLK) high energy evolution equations~\cite{Balitsky:2006wa,Balitsky:2008zza,Balitsky:2009xg,Balitsky:2013fea,Kovner:2013ona,Lappi:2015fma,Iancu:2015vea,Iancu:2015joa,Lappi:2016fmu}, and DIS cross sections~\cite{Balitsky:2010ze,Beuf:2011xd}.

An important advance in pushing the CGC framework to NLO accuracy was the calculation \cite{Chirilli:2011km,Chirilli:2012jd}
of NLO  single inclusive  particle production  in forward proton-nucleus collisions (see also the earlier works \cite{Dumitru:2005gt,Altinoluk:2011qy,JalilianMarian:2011dt}). 
We will here refer to the cross section formulae derived in Refs.~\cite{Chirilli:2011km,Chirilli:2012jd} as the ``CXY'' result according to the authors.
Here it was shown that the divergences in the rapidity (or longitudinal momentum) and transverse momentum integrals appearing in the NLO calculation can be factorized into the BK and DGLAP evolution of the target and projectile, respectively. In a  subsequent calculation~\cite{Stasto:2014sea} it was shown that in the large transverse momentum limit the calculation reduces to the appropriate tree-level process in collinear factorization, although in this case without a factorization of the rapidity divergence.

In the first numerical implementation~\cite{Stasto:2013cha} of the factorization framework of~\cite{Chirilli:2011km,Chirilli:2012jd} the NLO corrections turned out to be large and negative at large transverse momenta of the produced particles, to the extent that the total cross section becomes negative. A solution to this problematic behavior has been suggested to lie in the detailed implementation of the factorization of the rapidity divergence~\cite{Kang:2014lha} or in a kinematical constraint that must be imposed on the phase space of emitted gluons~\cite{Altinoluk:2014eka}. Indeed a recent implementation~\cite{Watanabe:2015tja} of this phase space constraint has alleviated the problem, without however removing it completely.

In this paper, we suggest an alternative way of implementing BK-factorization in the calculation of Chirilli et al.~\cite{Chirilli:2011km,Chirilli:2012jd} that combines aspects of these previous works. Our suggestion includes, as in~\cite{Kang:2014lha}, an explicit rapidity factorization scale in the ``hard functions'' of the NLO part of the cross section. Since the dependence of the cross section on this scale cancels against the rapidity scale to which the target is evolved, the total cross section formulation is explicitly independent of this factorization scale. Here we suggest implementing the ``kinematical constraint'' or, more precisely, ordering in light cone energy, by making this factorization scale dependent on the transverse momentum of the produced particle. We show that this allows for a renormalization prescription that makes the negative large momentum contribution to the cross section arbitrarily small. Combined with a corresponding form of the BK equation this should in the future make it possible to resum the problematic contributions at large transverse momentum similarly as recently suggested for the NLO BK equation~\cite{Iancu:2015vea,Lappi:2016fmu}.

For simplicity we will here only address the quark channel $q \to q$ and perform numerical calculations only for the Golec-Biernat and Wüsthoff (GBW)~\cite{GolecBiernat:1998js} parametrization of the dipole cross section. The gaussian $\ktt$-spectrum at leading order in the GBW model has the advantage for the purpose of this paper of being very clearly distinct from the power-law behavior of the NLO contributions. This exacerbates the problem of negative cross sections and should therefore put any attempt to stabilize the perturbative expansion to a more stringent test. A fuller phenomenological analysis of single inclusive particle production would require implementing also the gluon initiated channel, and a more realistic BK-evolved dipole cross section. Ultimately this should include a solution of the NLO BK equation~\cite{Iancu:2015vea,Lappi:2016fmu} and a simultaneous NLO fit to DIS data. At present we leave these further steps for future work.

This paper is structured as follows. We will first, in Sec.~\ref{sec:nlosinc} briefly recall CXY results~\cite{Chirilli:2011km,Chirilli:2012jd} for the NLO corrections to the cross section in the quark-initiated channel and how the factorization into DGLAP and BK evolution is performed there. We will then discuss the introduction of an explicit rapidity factorization scale as advocated in~\cite{Kang:2014lha} and explicitly show the modification to the NLO spectrum resulting from a variation of this scale in Sec.~\ref{sec:rapscale}. We then, in Sec.~\ref{sec:kc}, discuss imposing the  additional ``kinematical'' constraint of $k^-$-ordering (see e.g.~\cite{Beuf:2014uia}) on the rapidity factorization, which we implement as a momentum dependence of the rapidity factorization scale. Section \ref{sec:disc} then concludes with a brief discussion of possible future steps.

\section{Single inclusive particle production at NLO}
\label{sec:nlosinc}

Our starting point are the CXY formulae derived in Ref.~\cite{Chirilli:2012jd}. We will concentrate here on the quark channel, for which  we write the CXY result as:
\begin{widetext}
\begin{eqnarray}\label{eq:nlosigma}
 \frac{\ud N^{pA\to hX}}{\ud^2\pt \ud y_h}
&=&
\int_\tau^1 \frac{\ud z}{z^2}D_{h/q}(z) x_p q(x_p) \frac{\scal^{(0)}(\ktt) }{(2\pi)^2}
\\ \nonumber
&& +  \frac{\as}{2\pi^2} \int \frac{\ud z}{z^2}D_{h/q}(z) 
\int_{\tau/z}^1 \ud \xi \frac{1+\xi^2}{1-\xi}
\frac{x_p}{\xi} q\left(\frac{x_p}{\xi}\right) \left\{\cf \ical(\ktt,\xi) + \frac{\nc}{2}\jcal(\ktt,\xi) \right\}
\\ \nonumber
&& - \frac{\as}{2\pi^2} \int \frac{\ud z}{z^2}D_{h/q}(z) 
\int_{0}^1 \ud \xi \frac{1+\xi^2}{1-\xi}
x_p q\left(x_p \right) \left\{\cf \ical_v(\ktt,\xi) + \frac{\nc}{2}\jcal_v(\ktt,\xi) \right\},
\end{eqnarray}
where 
\begin{eqnarray}
\ical(\ktt,\xi) &=&
\int \frac{\ud^2 \qt}{(2\pi)^2} \scal(\qtt)
\left[\frac{\kt-\qt}{(\kt-\qt)^2} - \frac{\kt-\xi \qt}{(\kt-\xi \qt)^2} \right]^2
\\ 
% ******************
\jcal(\ktt,\xi) &=&
\int \frac{\ud^2 \qt}{(2\pi)^2} 
\frac{2(\kt-\xi\qt)\cdot(\kt-\qt)}{(\kt-\xi\qt)^2(\kt-\qt)^2}
\scal(\qtt)
-\int \frac{\ud^2 \qt}{(2\pi)^2} \frac{ \ud^2\lt}{(2\pi)^2}
\frac{2(\kt-\xi\qt)\cdot(\kt-\lt)}{(\kt-\xi\qt)^2(\kt-\lt)^2}
\scal(\qtt)\scal(\ltt)
\\
% ******************
\ical_v(\ktt,\xi) &=&
\scal(\ktt)\int \frac{ \ud^2 \qt }{(2\pi)^2}
\left[\frac{\kt-\qt}{(\kt-\qt)^2} - \frac{\xi\kt-\qt}{(\xi \kt-\qt)^2} \right]^2
\\
% ******************
\jcal_v(\ktt,\xi) &=&
\scal(\ktt)
\left[
\int \frac{\ud^2 \qt}{(2\pi)^2} 
\frac{2(\xi\kt-\qt)\cdot(\kt-\qt)}{(\xi\kt-\qt)^2(\kt-\qt)^2}
-\int \frac{\ud^2 \qt}{(2\pi)^2} \frac{  \ud^2\lt}{(2\pi)^2}
\frac{2(\xi\kt-\qt)\cdot(\lt-\qt)}{(\xi\kt-\qt)^2(\lt-\qt)^2}
\scal(\ltt)
\right].
\end{eqnarray}
\end{widetext}
Here we have  slightly altered the notation of \cite{Chirilli:2012jd} by including transverse momentum integrals in the functions $\ical,\jcal$ and leaving out an overall integration over the impact parameter $\bt$, thus our expression is for the multiplicity and not the cross section. The kinematical variables are defined as
$\pt = z\kt$, $x_p = \ptt e^{y_h}/(z\sqrt{s})$, $\tau = z x_p$, $x_g = \ptt/(z\sqrt{s})e^{-y_h}$, $\ptt=|\pt|$, $\qtt=|\qt|$, $\ktt=|\kt|$, and $\ltt=|\lt|$. Most important for our discussion here is the momentum fraction $\xi$: the fragmenting quark carries a fraction $\xi$ of the incoming quark longitudinal momentum. Thus the incoming quark has a momentum fraction $x_p/\xi$ of the incoming proton, where $x_p$ is the probe momentum fraction in leading order kinematics.  The radiated gluon in the NLO terms carries a longitudinal momentum fraction $1-\xi$: i.e. the limit $\xi\to 1 $ corresponds to the soft gluon emission that  must be resummed into BK evolution of the target.

These cross sections are all expressed in terms of the Fourier-transform of the fundamental representation dipole operator
\begin{eqnarray}
\scal(\ktt)=\scal(\ktt,\bt)=\int \ud^2\rt e^{-i\kt\cdot\rt} S(\rt),
\\ 
S(\rt=\xt-\yt)=\left< \frac{1}{\nc}\tr U(\xt)U^\dag(\yt) \right>.
\end{eqnarray}
 The dipole  is related to the notation of \cite{Chirilli:2012jd} by an overall integration over the impact parameter
\begin{equation}
 \fcal(\ktt) = \int\frac{\ud^2\bt}{(2\pi)^2} \scal(\ktt,\bt).
\end{equation}
In the following we leave out the explicit impact parameter dependence of $\scal(\ktt,\bt)$ from our notation.
Due to the unitarity of the Wilson lines $U$ the dipole cross section satisfies the normalization condition 
\begin{equation}\label{eq:sumrule}
 \int \frac{\ud^2\kt}{(2\pi)^2} \scal(\ktt) =1 .
\end{equation}
The expressions here use the mean field approximation, replacing 
the expectation value of the product of two dipole operators by the product of two expectation values $\scal(\qtt)\scal(\ltt)$. The superscript $(0)$ refers to the fact that at this stage the dipole operator in the leading order part of \nr{eq:nlosigma} is the unrenormalized ``bare'' dipole operator.

As noted in~\cite{Chirilli:2012jd}, several important features are already visible in these expressions. First, the terms with an explicit coefficient $\cf$
(i.e. $\ical$ and $\ical_v$) vanish in the limit $\xi \to 1 $ but have collinear divergences in the transverse momentum integration. These are nicely treated in~\cite{Chirilli:2012jd} by dimensional regularization in the transverse momentum integrals and factorized into the DGLAP evolution of the quark distribution function $q(x)$ and the fragmentation function $D_{h/q}(z)$. For these ``$\cf$-terms'' we will here follow the treatment of~\cite{Chirilli:2012jd}.
The terms with a coefficient $\nc/2$ (we will denote these as the ``$\nc$-terms'' in the following), i.e. $\jcal$ and $\jcal_v$, have  finite transverse integrals\footnote{To see this for the virtual term $\jcal_v$, one has to note that at large $\qt$ one can use the normalization \nr{eq:sumrule} to perform the $\lt$ integral, leading to a cancellation between the UV-divergences in the two terms of $\jcal_v$.} but are finite in the limit $\xi\to 1$.  Thus they produce a rapidity divergence due to the explicit factor $1/(1-\xi)$ in the expression for the multiplicity. If one takes the large $\nc$-limit, there are additional cancellations between some $\cf$ and $\nc$-terms that are used in~\cite{Chirilli:2012jd}; we will, however, not take this limit here. 

\section{Choosing the rapidity renormalization scale}
\label{sec:rapscale}

\subsection{Explicit subtraction scale}
In the CXY calculation the rapidity divergence is subtracted  by defining a renormalized dipole cross section as
\begin{multline}\label{eq:cxysub}
\scal(\ktt)  =  \scal^{(0)}(\ktt)
\\ + 2 \as \nc  \int_0^1 \frac{\ud \xi}{1-\xi} 
\left[\jcal(\ktt,1) - \jcal_v(\ktt,1)\right] .
\end{multline}
Expressed in coordinate space this reduces to a more familiar looking form in terms of an integral form of the BK renormalization group equation~\cite{Balitsky:1995ub,Kovchegov:1999yj}
\begin{multline}
S(\xt-\yt) = S^{(0)}(\xt-\yt)
- \frac{ \as \nc}{2\pi^2} \int_0^1 \frac{\ud \xi}{1-\xi}
 \int \ud^2 \zt
\\
\rule{0pt}{5ex}\times \frac{(\xt-\yt)^2}{(\xt-\zt)^2(\yt-\zt)^2}
\left[ S(\xt-\yt) - S(\xt-\zt)S(\zt-\yt) \right] .
\end{multline}
As pointed out in~\cite{Kang:2014lha}, this subtraction really should include an explicit rapidity factorization scale. The remaining NLO cross section would then depend on this factorization scale, which cancels at this order in perturbation theory against the rapidity up to which the dipole cross section in the leading order cross section is evolved. This is completely analogous with the subtraction of the collinear divergences into DGLAP evolution in the $\cf$-terms, which leaves the NLO cross sections explicitly dependent on a transverse momentum factorization scale.
Let us now accordingly include the rapidity factorization scale and subtract the rapidity divergence as 
\begin{multline}\label{eq:xifsub}
\scal(\ktt)  =  \scal^{(0)}(\ktt)
\\
+ 2 \as \nc \int\limits_{\xif}^1 \frac{\ud \xi}{1-\xi} 
\left[\jcal(\ktt,1) - \jcal_v(\ktt,1)\right] .
\end{multline}
The original choice of~\cite{Chirilli:2012jd} simply corresponds to $\xif=0$ in our notation. With this subtraction, and the factorization of the transverse divergence into DGLAP evolution, the cross section now becomes
\begin{widetext}
\begin{eqnarray}\label{eq:subtracted1}
 \frac{\ud N^{pA\to hX}}{\ud^2\pt \ud y_h}
&=&
\int_\tau^1 \frac{\ud z}{z^2}D_{h/q}(z) x_p q(x_p) \frac{S(\ktt) }{(2\pi)^2} + \cf \frac{\as}{2\pi^2} \int \frac{\ud z}{z^2}D_{h/q}(z)
\int_{\tau/z}^1 \ud \xi \frac{x_p}{\xi} q\left(\frac{x_p}{\xi}\right) \ical_\text{finite}(\ktt,\xi) \\ \nonumber
&& + \frac{\nc}{2} \frac{\as}{2\pi^2} \int \frac{\ud z}{z^2}D_{h/q}(z) \left\{
\int_{\tau/z}^{\xif} \ud \xi \frac{1+\xi^2}{1-\xi}
\frac{x_p}{\xi} q\left(\frac{x_p}{\xi}\right) \jcal(\ktt,\xi) - \int_{0}^{\xif} \ud \xi \frac{1+\xi^2}{1-\xi}
x_p q\left(x_p \right)
\jcal_v(\ktt,\xi) \right. \\ \nonumber
&& \hspace{4cm}+ \left.\int_{\xif}^{1} \ud \xi \frac{1}{1-\xi}
\left[\kcal(\xi)-\kcal(1)\right] \right\} ,
\end{eqnarray}
with
\begin{eqnarray}
\ical_\text{finite}(\ktt,\xi) &=& \pi \int \frac{\ud^2\rt}{(2\pi)^2} S(\rt) \left[ \mathcal{P}_{qq}(\xi) \ln \frac{c_{0}^{2}}{r_{\perp }^{2}\mu ^{2}}\left( e^{-ik_{\perp }\cdot r_{\perp }}+\frac{1}{\xi
^{2}}e^{-i\frac{k_{\perp }}{\xi }\cdot r_{\perp }}\right) -3 \delta
(1-\xi ) e^{-ik_{\perp }\cdot r_{\perp }}\ln \frac{c_{0}^{2}}{r_{\perp
}^{2}k_{\perp }^{2}} \right]   \nonumber \\
&& + \frac{\scal(\ktt)}{2\pi} \left(\frac{(1+\xi^2)\ln(1-\xi)^2}{1-\xi}\right)_+ - 2\frac{1+\xi^2}{(1-\xi)_+} I_{2 1}(\ktt,\xi),
\end{eqnarray}
\end{widetext}
where $\mathcal{P}_{qq}$ is the quark-quark splitting function
\begin{equation}
\mathcal{P}_{qq}(\xi)=\left(\frac{1+\xi^2}{1-\xi}\right)_+,
\end{equation}
and the functions $I_{2 1}(\ktt,\xi)$ and $\kcal(\xi)$ are defined according to
\begin{align}
& I_{2 1}(\ktt,\xi) = \int \frac{\ud^2 \qt}{(2\pi)^2} \frac{(\kt-\xi\qt)\cdot(\kt-\qt)}{(\kt-\xi\qt)^2(\kt-\qt)^2} \scal(\qtt) \nonumber \\
& \hspace{1.8cm} + \frac{1}{4\pi}\scal(\ktt)\ln(1-\xi)^2, \\
& \kcal(\xi)\!=\!(1\!+\!\xi^2) \bigg[\frac{x_p}{\xi} q\left(\!\frac{x_p}{\xi}\!\right) \jcal(\ktt,\xi)\!-\!x_p q\left(x_p \right)
\jcal_v(\ktt,\xi) \bigg].
\end{align}

In these expressions we used the plus prescription
\begin{eqnarray}
\int_{x_0}^1 \ud x \left(f(x)\right)_+ g(x)&=&\int_{x_0}^1 \ud x f(x)\left[g(x)-g(1)\right] \nonumber \\
&&-g(1)\int^{x_0}_0 \ud x f(x) \, ,
\end{eqnarray}
where $f(x)$ is singular at $x=1$ and $g(x)$ is a regular function.

Now we can see that if $\xif$ is chosen to be very close to one (i.e. if one makes sure to subtract only terms that have a very large energy logarithm), the contribution of the last $\nc$-term (integrated from $\xif$ to 1) is negligible $\sim (1-\xif)$.  The first two $\nc$-terms, on the other hand, yield a large logarithmic contribution $\sim \ln (1-\xif)$ from the upper limit of the integration.

\subsection{Analytical considerations}
\label{sec:analytical}

It is instructive to see how these expressions for the subtracted cross section behave in the opposite limits of  small and large transverse momentum for the produced quark. At large transverse momentum the result can be easily obtained from \cite{Stasto:2014sea}, where the unsubtracted cross sections have been matched to collinear perturbation theory. In the large $\ktt$ limit only the radiative corrections contribute and the leading behavior comes from
\begin{eqnarray}
 \ical(\ktt,\xi) &\approx& \frac{(1-\xi)^2}{\ktt^4} \frac{\alpha_s 2 \pi^2}{\nc\stt} xG(x,\mu),\label{eq:largekt-CF}
\\
 \jcal(\ktt,\xi) &\approx& \frac{2\xi}{\ktt^4}\frac{\alpha_s 2 \pi^2}{\nc\stt} xG(x,\mu), \label{eq:largekt-Nc}
\end{eqnarray}
where the integrated gluon distribution is
\begin{equation}
 xG(x,\mu) = \frac{\nc\stt}{\alpha_s 2 \pi^2} \int \frac{\ud^2\qt}{(2\pi)^2} \qt^2 \scal(\qtt),
\end{equation}
with $\stt$ the transverse area of the target hadron.

It is easy to see that if we replace the expression of $\ical$ in Eq.~(\ref{eq:nlosigma}) by its large $\ktt$ limit~(\ref{eq:largekt-CF}), the $\cf$-term will yield a positive contribution. On the other hand, the large $\ktt$ behavior of $\jcal$~(\ref{eq:largekt-Nc}) means that the subtracted $\nc$-term in Eq.~(\ref{eq:subtracted1}),
\begin{equation}
\int_{\xif}^{1} \ud \xi \frac{1}{1-\xi}
\left[\kcal(\xi)-\kcal(1)\right],
\end{equation}
will yield a contribution that behaves as a power in $\ktt$ and is negative due to the growing $\xi$-dependence of $\kcal$ in this limit.
For values of $\xif$ close to 0, the magnitude of this negative term is larger than the one of the (positive) $\cf$-term. Therefore, if the leading order cross section falls rapidly at large $\ktt$ (as in the GBW parametrization), the leading large $\ktt$ behavior comes from the NLO corrections and the cross section is negative.
Thus the negativity of the NLO cross section can be traced  back to the fact that the unsubtracted cross section at large $\ktt$ is proportional to $\xi$.   Subtracting, as in CXY (see \nr{eq:cxysub}), the integral over the whole $\xi$-interval multiplied by the cross section at $\xi=1$ subtracts a very large finite contribution in addition to the divergent part. \emph{This oversubtraction of the rapidity divergence is what makes the cross section negative at large transverse momenta.} This was the main point in~\cite{Kang:2014lha}, where the authors showed that reintroducing a part of this oversubtracted contribution again makes the cross section positive. When we increase $\xif$ towards 1, the large negative contribution dies away.

For practical purposes, this is not a formalism that we would wish to use for extremely small transverse momenta, since the independent vacuum fragmentation picture of hadron production is probably not a valid physical picture there. However, from a formal point of view it can be instructive to see what happens in the small $\ktt$ limit.
The fact that BK evolution preserves the dilute limit $S(r)\to 1$ when $r \to 0$, i.e. the sum rule \nr{eq:sumrule}, tells us that the quantity $(\jcal(\ktt,1)-\jcal_v(\ktt,1))$, which is positive in the large $\ktt$ limit, must be negative in some other regions of phase space. Indeed, if we denote by $\delta \scal(\ktt)$ the contribution that is added to $\scal(\ktt)$ in one rapidity step of BK evolution, the sum rule ensures that the $\kt$-integral of $\delta \scal(\ktt)$ is zero.
At large $\ktt$, $(\jcal(\ktt,1)-\jcal_v(\ktt,1))$ is positive. Therefore, if we increase $\xif$ we subtract a smaller but positive contribution and thus the cross section increases. On the other hand, at smaller $\ktt$ we expect $(\jcal(\ktt,1)-\jcal_v(\ktt,1))$ to be negative and the cross section to decrease with increasing $\xif$.

As both $\jcal$ and $\jcal_v$ are free of IR and UV divergences, we can estimate their leading order behavior by using 
\begin{equation}
 \int_0^{\qt_0} \frac{\ud^2 \qt}{(2\pi)^2} 
\frac{2(\lt-\qt)\cdot(\kt-\qt)}{(\lt-\qt)^2(\kt-\qt)^2}= \frac{1}{2\pi} \ln\frac{\qt_0^2}{(\lt-\kt)^2} 
\end{equation}
with $ |\qt_0| > \ktt, |\qt_0| > \ltt$, which allows us to figure out the small $\ktt$ limit of $\jcal$ and $\jcal_v$ as 
\begin{align}
\label{eq:jsmallk}
\jcal(\ktt,\xi) & \approx \frac{\scal({\qtt}_0)}{2\pi\xi} \ln\frac{Q^2 \xi²}{\kt^2(1-\xi)^2} , \\
\label{eq:jvsmallk}
\jcal_v(\ktt,\xi) & = \frac{\scal(\ktt)}{2\pi} \int \frac{\ud^2 \lt}{(2\pi)^2}\scal(\ltt)\ln\frac{(\lt-\xi \kt)^2}{\kt^2(1-\xi)^2}\nonumber \\
& \approx \frac{\scal(\ktt)}{2\pi} \ln\frac{Q'^{2}}{\ktt^2(1-\xi)^2}.
\end{align}
Here $Q$ and $Q'$ are some hard momentum scales which are much larger than $\ktt$ and ${\qtt}_0$ is chosen such that the integral of  $\scal(\qtt) f(\qtt)$ equals to $\scal({\qtt}_0)[ F(Q)-F(0)]$ with $F(\qtt)$ the integral of the function $f(\qtt)$. From the above expressions, we see that both $\jcal$ and $\jcal_v$ have logarithmic divergences at $\xi =1$ and $\ktt=0$. It is known that
$(\jcal(\ktt,\xi)-\jcal_v(\ktt,\xi))$ should be finite at $\xi=1$, and thus the $\ln(1-\xi)^2$ terms in Eqs.~\nr{eq:jsmallk} and~\nr{eq:jvsmallk} must have the same coefficient. Therefore $\scal({\qtt}_0)$ has to be $\scal({\ktt})$ at $\xi=1$. This also tells us that,  at $\xi=1$,  $\ln\ktt^2$ cancels out between the real and virtual terms. Clearly, there is a $\ln\ktt^2$ behavior in $(\jcal(\ktt,\xi)-\jcal_v(\ktt,\xi))$ when $\xi\neq 1$.

In the GBW model it is possible to study analytically the behavior of the subtraction term. In this model the dipole cross section is given by
\begin{equation}
S(\rt)=e^{-\rt^2 \qs^2/4}
\end{equation}
which leads to
\begin{equation}
\scal(\ktt)=\frac{4\pi}{\qs^2}e^{-\ktt^2/\qs^2}
\end{equation}
For simplicity we will consider here that the saturation scale $\qs$ is a constant. In this model the real $\nc$ term reads
\begin{widetext}
\begin{equation}\label{eq:jcal_GBW}
\jcal(\ktt,\xi)=2\left[I_{2 1}(\ktt,\xi)-\frac{e^{-\ktt^2/\qs^2} }{\qs^2}\ln(1-\xi)^2-\frac{1 }{\ktt^2}\left(1-e^{-\ktt^2/\qs^2}\right) \left(1-e^{-\ktt^2/(\xi^2 \qs^2)}\right)\right] ,
\end{equation}
with
\begin{equation}
I_{2 1}(\ktt,\xi)=\frac{e^{-\ktt^2/(\xi \qs^2)}}{\xi \qs^2} \left[\text{Ei}\left(\frac{\ktt^2}{\xi \qs^2}\right)-\text{Ei}\left(\frac{\ktt^2 (\xi-1)}{\xi^2 \qs^2}\right)-\text{Ei}\left(\frac{\ktt^2 (1-\xi )}{\xi \qs^2}\right)\right]+\frac{e^{-\ktt^2/\qs^2}}{\qs^2} \ln(1-\xi)^2 ,
\end{equation}
where $\text{Ei}(x)$ is the exponential integral function, $\text{Ei}(x)=-\int_{-x}^{\infty}\ud t \, \frac{e^{-t}}{t}$,
and the virtual term reads
\begin{equation}\label{eq:jcalv_GBW}
\jcal_v(\ktt,\xi)=2\frac{e^{-\ktt^2/\qs^2} }{\qs^2}\left[\Gamma\left(0,\frac{\ktt^2 \xi ^2}{\qs^2}\right)+\ln{\xi^2}-\ln(1-\xi)^2\right] .
\end{equation}
Thus the subtraction term is
\begin{equation}
	\jcal(\ktt,1)-\jcal_v(\ktt,1)=4\frac{e^{-\ktt^2/\qs^2}}{\qs^2} \left[-\gamma_E+\frac{\qs^2}{\ktt^2}-\frac{\qs^2}{\ktt^2} \cosh \left(\frac{\ktt^2}{\qs^2}\right)+ \frac{\text{Ei}\left(\ktt^2/\qs^2\right)+\text{Ei}\left(-\ktt^2/\qs^2\right)}{2}+\ln\left(\frac{\qs^2}{\ktt^2}\right)\right] .
\end{equation}
\end{widetext}
With this expression one can explicitly show that 
\begin{equation}
\int \ud^2\kt \left[\jcal(\ktt,1)-\jcal_v(\ktt,1)\right]=0 \, ,
\end{equation}
as required to satisfy the sum rule~(\ref{eq:sumrule}).

\begin{figure}
	\centering
	\includegraphics[width=\linewidth]{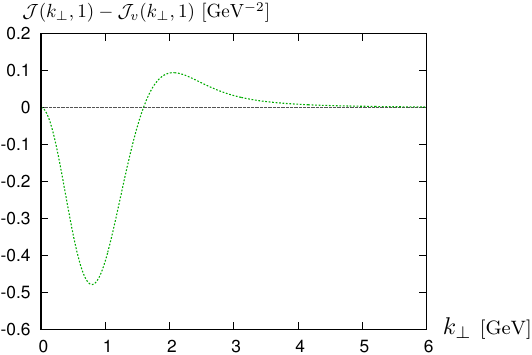}
	\caption{The subtraction term $\jcal(\ktt,1)-\jcal_v(\ktt,1)$ as a function of $\ktt$ in the GBW model for $\qs=1$ GeV.}
	\label{fig:subtraction}
\end{figure}

At small $\ktt$, $\jcal$ and $\jcal_v$ read
\begin{align}
\jcal(\ktt,\xi) & \approx \frac{2}{\qs^2} \left(-\frac{\gamma_E}{\xi}+\frac{1}{\xi}\ln\frac{\qs^2\xi^2}{\ktt^2(1-\xi)^2 } -\frac{\ktt^2}{\qs^2} \right) , \\
\jcal_v(\ktt,\xi) & \approx \frac{2}{\qs^2}\left( -\gamma_E + \ln\frac{\qs^2}{\ktt^2(1-\xi)^2} \right)  ,
\end{align}
and the subtraction term behaves like
\begin{equation}\label{eq:sub_smallkt}
\jcal(\ktt,1)-\jcal_v(\ktt,1)\approx-\frac{2 \ktt^2}{\qs^4} \, .
\end{equation}
In Fig.~\ref{fig:subtraction}, we show the behavior of $\jcal(\ktt,1)-\jcal_v(\ktt,1)$ as a function of $\ktt$ for a fixed saturation scale $\qs=1$ GeV. We observe that, as expected from the general limit~(\ref{eq:largekt-Nc}), the subtraction term is positive at large $\ktt$. The precise functional form $\sim \ktt^2$ of the small $\ktt$ behavior, on the other hand, is specific to the GBW model.
This confirms explicitly that when we increase $\xif$,  we are subtracting less of the  positive contribution in the large $\ktt$ region and less of the negative contribution at small $\ktt$. Therefore we would expect that the subtracted multiplicity will increase at large $\ktt$ and decrease at small $\ktt$ in the GBW model with the increasing of $\xif$.

\subsection{Numerical results in the GBW model}

In this section we illustrate the features we have discussed so far using the GBW model to parametrize the dipole cross section $S(\rt)$. Again we focus on the contribution of the quark channel $q \to q$.
In this model, the saturation scale $\qs$ at a given $x$ is parametrized by
\begin{equation}
\qs^2=c A^{1/3} Q_{s 0}^2 \left(\frac{x_0}{x}\right)^{\lambda},
\end{equation}
where $A$ is the atomic number of the target and the values of the other parameters are $c=0.56$, $Q_{s 0}=1$ GeV, $x_0=3.04 \times 10^{-4}$, $\lambda=0.288$~\cite{GolecBiernat:1998js}. In the following the $x$ value at which we evaluate $\qs$ is given by $x=x_g/(1-\xif)$, where $x_g$ corresponds to the leading order kinematics: $x_g=\frac{\ptt}{z \sqrt{s}}e^{-y_h}$.
Since in the GBW model the dipole cross section has a simple gaussian form, it is possible to perform some integrals analytically. This avoids the need to deal with several oscillatory integrals in the numerical implementation which is thus simpler. It also enhances the difference in the behaviors of the leading and next-to-leading contributions at large $k_T$. Indeed, the leading order contribution is proportional to the Fourier transform of $S(\rt)$, which is also gaussian, while at large $k_T$ the NLO corrections have a power law behavior, as can be seen from Eqs.~(\ref{eq:largekt-CF}) and~(\ref{eq:largekt-Nc}).

The expression for $\ud \sigma^{pA\to hX}/\ud^2\pt \ud y_h$ in the GBW model can be found in Ref.~\cite{Chirilli:2012jd} in the large $\nc$ limit. Here we need to keep $\nc$ finite because we want to have a clear separation between the $\cf$-terms, which are associated with the collinear divergence, and the $\nc$-terms which are associated with the rapidity divergence. In this case, the multiplicity reads
\begin{widetext}
\begin{align}
\frac{\ud N^{pA\to hX}}{\ud^2\pt \ud y_h}
=&
\int\limits_\tau^1 \frac{\ud z}{z^2}D_{h/q}(z) x_p q(x_p) \frac{\scal(\ktt)}{(2\pi)^2}
\nonumber \\
& \hspace{-2.1cm} + \cf \frac{\as}{2\pi^2} \!\! \int \! \frac{\ud z}{z^2}D_{h/q}(z) \!\!
\int\limits_{\tau/z}^1 \!\! \ud \xi \frac{x_p}{\xi}q\!\left(\frac{x_p}{\xi}\right) \! \frac{1}{4\pi} \Bigg\{\!
%=====================================
\mathcal{P}_{qq}(\xi ) \scal(\ktt) \left[ \ln \frac{\qs^2}{\mu^2 e^{\gamma_E}} +\mathcal{L}\left(-\frac{\ktt^2}{\qs^2}\right)\right] + 2 \scal(\ktt) \left( \frac{\left( 1+\xi ^{2}\right) \ln \left( 1-\xi \right) ^{2}}{1-\xi }\right)_{\!\! +} \nonumber \\
& \hspace{4.7cm} + \! \frac{1}{\xi^2}\mathcal{P}_{qq}(\xi) \scal\left(\frac{\ktt}{\xi}\right) \! \left[ \ln \frac{\qs^2}{\mu^2 e^{\gamma_E}} + \mathcal{L}\left(-\frac{\ktt^2}{\xi^2 \qs^2}\right)\right] - 8\pi \frac{1+\xi^2}{(1-\xi)_+} I_{2 1}(\ktt,\xi) \nonumber \\
& \hspace{4.7cm} - 3 \delta(1-\xi) \scal(\ktt) \left[ \ln \frac{\qs^2}{\ktt^2 e^{\gamma_E}} + \mathcal{L}\left(-\frac{\ktt^2}{\qs^2}\right)\right] \Bigg\} \\
%=======================================
& \hspace{-2.1cm} + \frac{\nc}{2} \frac{\as}{2\pi^2} \!\! \int \! \frac{\!\ud z}{z^2}D_{h/q}(z) \left\{\!
\int_{\tau/z}^{\xif} \!\! \ud \xi \frac{1+\xi^2}{1-\xi}
\frac{x_p}{\xi} q\left(\!\frac{x_p}{\xi}\!\right) \! \jcal(\ktt,\xi) \!-\! \int_{0}^{\xif} \!\! \ud \xi \frac{1+\xi^2}{1-\xi}
x_p q\left(x_p \right)
\jcal_v(\ktt,\xi) \!+\! \int_{\xif}^{1} \!\! \ud \xi \frac{1}{1-\xi}
\left[\kcal(\xi)\!-\!\kcal(1)\right] \right\} \nonumber 
\end{align}
\end{widetext}
where
\begin{equation}
\mathcal{L}(x) = -\gamma_E-\Gamma(0,x)-\ln{x} \, ,
\end{equation}
and the expressions for $\jcal$ and $\jcal_v$ in the GBW model can be read from Eqs.~(\ref{eq:jcal_GBW}) and~(\ref{eq:jcalv_GBW}), respectively.
Besides the absence of the impact parameter integration, there are two differences with the corresponding expressions given in Ref.~\cite{Chirilli:2012jd}: first, we keep terms proportional to $\cf-\nc/2$ which vanish in the large $\nc$ limit taken in Ref.~\cite{Chirilli:2012jd}. Second, we modify the rapidity divergence subtraction by using the cutoff $\xif$ introduced previously.

\begin{figure}
	\centering
	\includegraphics[width=\linewidth]{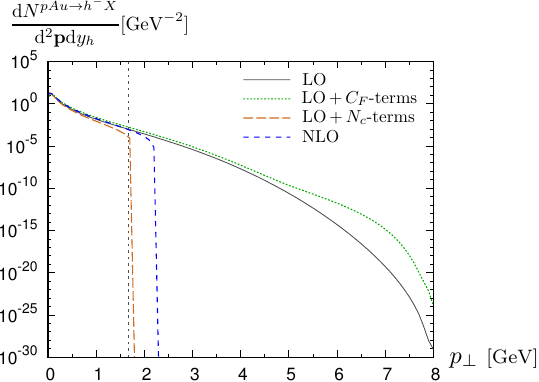}
	\caption{Multiplicity as a function of $\ptt$ in the GBW model at leading and next-to-leading order and when including only the $\cf$ or $\nc$ NLO corrections. The vertical dashed line corresponds to $\qs \approx \ptt$.}
	\label{fig:dN_nocutoff}
\end{figure}

\begin{figure}
	\centering
	\includegraphics[width=\linewidth]{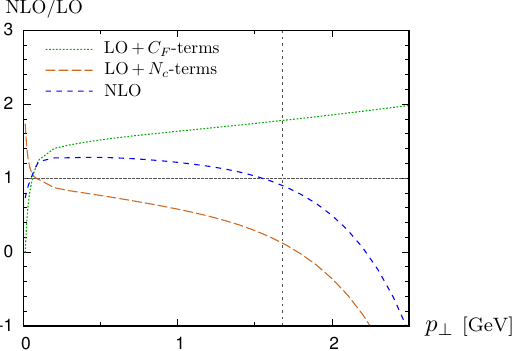}
	\caption{Ratio of the multiplicity at next-to-leading and leading order for $\xif=0$, when including only the $\cf$ or $\nc$ NLO corrections or both. The vertical dashed line corresponds to $\qs \approx \ptt$.}
	\label{fig:ratio_nocutoff}
\end{figure}

In Fig.~\ref{fig:dN_nocutoff} we show the multiplicity $\ud N^{p Au\to h^- X}/\ud^2\pt \ud y_h$ as a function of $\ptt$ in the GBW model for $\xif=0$ which corresponds to the choice made in Ref.~\cite{Chirilli:2012jd}. We take $\sqrt{s}=200$ GeV, $\alpha_s=0.2$, $\mu^2=10$ GeV$^2$ and $y_h=3.2$. For the collinear PDFs $q(x)$ and fragmentation functions $D_{h/q}(z)$ we use the MSTW 2008~\cite{Martin:2009iq} and DSS~\cite{deFlorian:2007aj} parametrizations, respectively, both at next-to-leading order.
As observed in Ref.~\cite{Stasto:2013cha}, when $\xif=0$ the NLO multiplicity is negative when $\ptt$ is larger than a certain value, of the order of the saturation scale. As discussed above, this negativity comes from the $\nc$-terms. This  can be seen from the same figure where we also show the effect of including only the NLO corrections proportional to $\cf$ or $\nc$. At large transverse momentum the $\cf$ corrections are positive while the $\nc$ corrections are negative and large enough to make the total NLO multiplicity negative. We observe that when including only the NLO contributions proportional to $\nc$ the multiplicity becomes negative very close to the point where $\ptt \approx \qs$.
To see more clearly the behavior of the NLO corrections at small transverse momentum, we show in Fig.~\ref{fig:ratio_nocutoff} the ratio of the NLO and LO multiplicity for $\ptt \le 2.5$ GeV when including only the $\cf$ or $\nc$ contributions or both. One can note that, already for values of $\ptt$ of the order of 2 GeV, both the $\cf$ and $\nc$ contributions are of the same order of magnitude as the LO term.

\begin{figure}
	\centering
	\includegraphics[width=\linewidth]{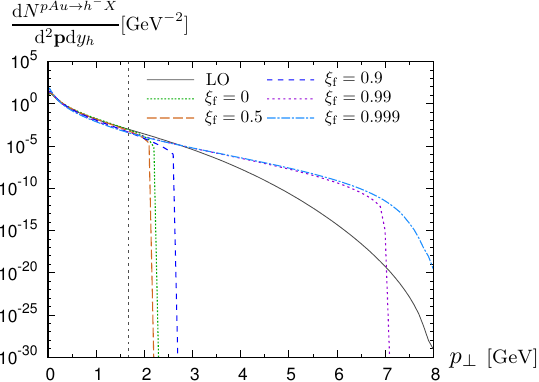}
	\caption{Multiplicity as a function of $\ptt$ in the GBW model at NLO for different values of $\xif$ compared with the LO result. The vertical dashed line corresponds to $\qs \approx \ptt$.}
	\label{fig:dN_cutoff}
\end{figure}

\begin{figure}
	\centering
	\includegraphics[width=\linewidth]{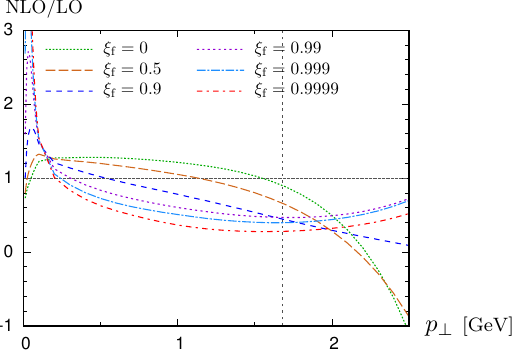}
	\caption{Ratio of the multiplicity at next-to-leading and leading order for different values of $\xif$. The vertical dashed line corresponds to $\qs \approx \ptt$.}
	\label{fig:ratio_cutoff}
\end{figure}

\begin{figure}
	\centering
	\includegraphics[width=\linewidth]{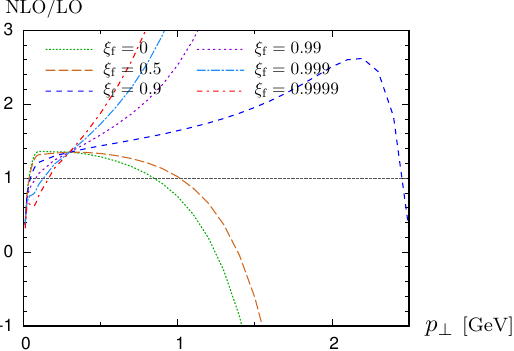}
	\caption{Ratio of the multiplicity at next-to-leading and leading order for different values of $\xif$ with $\qs=1$ GeV.}
		\label{fig:ratio_cutoff_Qs_1}
\end{figure}

As discussed in Sec.~\ref{sec:rapscale}, instead of subtracting the whole $\xi$ interval when subtracting the rapidity divergence~\nr{eq:cxysub}, one can introduce an explicit cutoff $\xif$ to determine which contributions are included in the renormalized dipole cross section~\nr{eq:xifsub}. We recall that, compared to the results shown above for $\xif=0$, only the $\nc$-terms are affected by this procedure. In Fig.~\ref{fig:dN_cutoff}, we show the multiplicity for several fixed values of $\xif$ as a function of $\ptt$. We observe that, for the reason exposed in Sec.~\ref{sec:analytical}, values of $\xif$ close to 1 lead to a positive multiplicity up to larger values of $\ptt$. In particular, for $\xif \gtrsim 0.999$, the multiplicity is positive up to $\ptt=8$ GeV. In Fig.~\ref{fig:ratio_cutoff}, we show the ratio NLO/LO for several values of $\xif$ and $\ptt \le 2.5$ GeV. Here we see that values of $\xif$ very close to 1, corresponding to a positive multiplicity at large $\ptt$, lead to a NLO multiplicity smaller than at leading order at moderate $\ptt$.
In Fig.~\ref{fig:ratio_cutoff_Qs_1}, we show the same ratio with a fixed value of the saturation scale, $\qs=1$ GeV. Here we see clearly that, as could be expected from the small $\ktt$ behavior of the subtraction term~(\ref{eq:sub_smallkt}), larger values of $\xif$ lead to smaller cross sections at small $\ptt$ and larger cross sections at large $\ptt$.

To conclude this section, it appears that the choice of the value of $\xif$ can have an important impact on the final results, both at small and large transverse momentum. In the following we propose a way to fix the value of this parameter based on physical considerations.

\section{Ordering in light cone energy}
\label{sec:kc}

It has recently been emphasized~\cite{Beuf:2014uia} that, in order to have a stable BK evolution beyond leading order, the gluon cascade resummed by the evolution should be \emph{ordered in the light cone energy $k^-$} of the projectile. The requirement of ordering in $k^-$ can equivalently be thought of as an ordering in its conjugate variable $x^+$, which is the light cone lifetime  or ``Ioffe time'' of fluctuations in the projectile~\cite{Altinoluk:2014eka}. In many works, this feature is known as the ``kinematical constraint'' or as imposing ``exact kinematics''~\cite{Motyka:2009gi,Stasto:2014sea,Watanabe:2015tja}, a terminology that we will comment on in Sec.\ref{sec:disc}. Imposing $k^-$-ordering has also been one ingredient in the program of ``small-$x$ resummation'' in the context of the linear BFKL evolution~\cite{Salam:1998tj,Ciafaloni:1999yw,Altarelli:1999vw,Ciafaloni:2003rd}. 
Since we are working in a frame where we consider the gluons as being emitted from the probe, the emitted gluon always has a smaller $k^+$ momentum than its parent quark; the emissions are therefore naturally ordered in $k^+$. The requirement of $k^-$ ordering, on the other hand,  must be imposed separately.

At leading order, the physical picture of the scattering is that of an incoming collinear quark (i.e. with only a $k^+$ momentum component)  that acquires a transverse momentum and light cone energy ($\kt$ and $k^-$) from the target. Considering the production of a quark with a fixed $\kt$, the light cone energy required is
\begin{equation}
 k^-_\text{LO} = \frac{\kt^2}{2k^+} = \frac{\kt^2}{2 x_p P^+}
\end{equation}
The  fraction of the target momentum required to set the produced quark on shell is defined by the leading order kinematics as $x_g = k^-_\text{LO}/P^-$, where $P^-$ is the momentum of the target. Thus we can write $1/(2 x_p P^+) = x_g P^- /\kt^2$. 

\begin{figure} 
\includegraphics[scale=1.1]{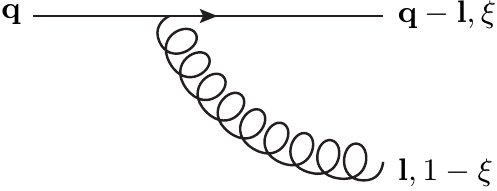}
\caption{Gluon emission}\label{fig:q2qg}
\end{figure}

We now want to implement $k^-$-ordering in the calculation of the single inclusive cross section. For this purpose let us consider the emission of a gluon from the incoming quark that is present in all of the contributing diagrams. Labeling the transverse momentum of the incoming quark as $\qt$ and radiated gluon as $\lt$ (cf. Fig.~\ref{fig:q2qg}), the light cone energy introduced from the gluon emission is
\begin{align}\label{eq:kminusqg}
\Delta k^-_{qg} &= 
\frac{1}{2x_pP^+} \left[ \frac{\lt^2}{1-\xi} + \frac{(\qt-\lt)^2}{\xi} - \qt^2\right] \nonumber \\ 
&=\frac{x_g P^-}{\kt^2} \frac{(\lt-(1-\xi)\qt)^2}{\xi(1-\xi)},
\end{align}
where $1-\xi$ is the momentum fraction of the emitted gluon.
In some diagrams, the splitting happens before the interaction with the target and $\qt=0$, in final state radiation diagrams $\qtt\sim \qs$. Similarly, for some diagrams, but not all, the quark does not interact with the target after the emission and $\qt-\lt=\kt$. In general, the momenta $\lt,\qt$ in \eq\nr{eq:kminusqg} are integrated over in the diagram, whereas the transverse momentum $\kt$ is that of the produced quark and is kept fixed (at a fixed fragmentation $z$). It appears as an overall prefactor in \nr{eq:kminusqg} when the incoming quark longitudinal momentum $x_pP^+$ is expressed in terms of the target variables $x_g$ and $P^-$ and is therefore common to all diagrams contributing to the cross section.

We now want to derive a renormalization group equation describing the target. This means that emissions in a certain kinematical regime have to be subtracted from the cross section and absorbed into a redefinition of the target, as was done in \eq\nr{eq:cxysub}. In particular, this kinematical regime to be subtracted should include the limit $\xi = 1$. Now the requirement of a $k^-$-ordering in the evolution means that the subtraction criterion should be a condition on $\Delta k^-$. Since $\Delta k^-\sim 1/(1-\xi)$ for $\xi\to 1$, this means that fluctuations around $\xi=1$ with $\Delta k^-$ larger than a certain factorization scale should be subtracted. Thus, our renormalization scheme is defined by the condition that all the contributions with
$ \xi \geq \xif(\ktt)$ are to be subtracted from the cross section. 
Most importantly, we want the subtracted region to include contributions with
\begin{equation}\label{eq:kminusorder}
\Delta k^-_{qg} = 
\frac{x_g P^-}{\kt^2} \frac{(\lt-(1-\xi)\qt)^2}{\xi(1-\xi)} \gtrsim \xf P^-,
\end{equation}
where $\xf$ is in principle an arbitrary factorization scale that will appear both in the leading order term as the rapidity up to which the target must be evolved, and in the hard factors of the NLO cross section. 

We emphasize that \nr{eq:kminusorder} is not a (momentum conservation) kinematical constraint for the process, but a renormalization condition specifying which parts of phase space should be subtracted from the cross section.
A natural choice that resums all the large energy logarithms into the evolution is to take $\xf \approx x_g$. In the typical kinematical regime when all the transverse momenta are of the same order, $\Delta k^-_{qg} \sim x_g/(1-\xi) \gtrsim \xf$ for all $\xi$, and we can safely subtract the whole $\xi$ interval as done in \cite{Chirilli:2011km,Chirilli:2012jd}. However, when the produced quark momentum $\kt$ is much larger than the typical target scale $\qs$, there are configurations where the $k^-$-ordering condition \nr{eq:kminusorder} is not satisfied for all $\xi$, because $\lt$ is integrated over in a range that includes typical target scales $\ltt \sim \qs$. Thus we should, for large $\kt$, only subtract values of $\xi$ close to 1 that satisfy
\begin{equation}
\Delta k^-_{qg} = 
\frac{x_g P^-}{\kt^2} \frac{\qs^2}{1-\xi} \geq \xf P^- ,
\end{equation}
or 
\begin{equation}
1-\xi\leq \frac{\qs^2}{\kt^2} \frac{x_g}{\xf} \sim \frac{\qs^2}{\kt^2}. 
\end{equation}
This line of argument leads us to our proposal to implement $k^-$-ordering or the ``kinematical constraint'' by  subtracting a $\ktt$-dependent fraction of the $\xi$ integral from the cross section as 
\begin{align}\label{eq:dlzsub}
\scal(\ktt) = \scal^{(0)}(\ktt)
+ 2 \as \nc \int_{\xif(\ktt)}^1 \frac{\ud \xi}{1-\xi} 
[\jcal(\ktt,1) \nonumber \\*
- \jcal_v(\ktt,1)],
\end{align}
with 
\begin{equation}\label{eq:xifmin}
\xif(\ktt) = 1 - \min \left\{\frac{x_g}{\xf} \frac{\qs^2}{\kt^2} ,1\right\}.
\end{equation}
Note that since there is a factorization scale $x_g/\xf$ in the limit, we have the freedom to change this scale with a factor of order 1. This change should cancel against a corresponding change in the rapidity to which the dipole amplitude is evolved.
An alternate, somewhat smoother form with the same parametric behavior would be
\begin{equation}\label{eq:xifcontinuous}
\xif(\ktt) = \frac{\ktt^2}{\ktt^2 + (x_g/\xf)\qs^2},
\end{equation}
where we can vary the factorization scale $x_g/\xf$ in an interval such as  $\half \dots 2$.

\begin{figure}
	\centering
	\includegraphics[width=\linewidth]{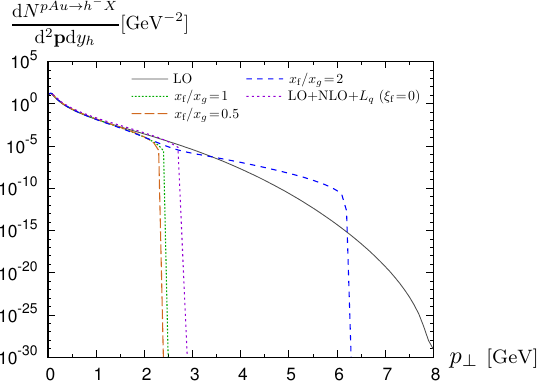}
	\caption{Multiplicity obtained using the parametrization~(\ref{eq:xifcontinuous}) of $\xif$ for different values of $\frac{\xf}{x_g}$ compared with the results at leading order and when using the correction term $L_q$ introduced in Ref.~\cite{Watanabe:2015tja}.}
	\label{fig:dN_xifcontinuous}
\end{figure}

\begin{figure}
	\centering
	\includegraphics[width=\linewidth]{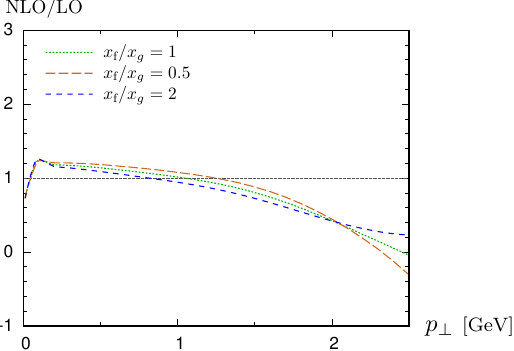}
	\caption{Ratio of the multiplicity at next-to-leading and leading order using the parametrization~(\ref{eq:xifcontinuous}) of $\xif$ for different values of $\frac{\xf}{x_g}$.}
	\label{fig:ratio_xifcontinuous}
\end{figure}

In practice, the two choices~(\ref{eq:xifmin}) and~(\ref{eq:xifcontinuous}) lead to very similar results. Therefore we will only show results obtained with the smoother choice~(\ref{eq:xifcontinuous}). In Figs.~\ref{fig:dN_xifcontinuous} and ~\ref{fig:ratio_xifcontinuous} we show the multiplicity as a function of $\ptt$ for three values of the ratio $x_g/\xf$. In the three cases, the NLO multiplicity still becomes negative for some value of the transverse momentum. For $x_g/\xf=1$ or 2 this happens approximately at the same point as for $\xif=0$, as can be seen by comparison with Fig.~\ref{fig:dN_nocutoff}. On the other hand, for $x_g/\xf=0.5$ the multiplicity is positive up to a significantly larger value of $\ptt$ of the order of 6~GeV. For comparison we also show in Fig.~\ref{fig:dN_xifcontinuous} the results obtained by following the approach of Ref.~\cite{Watanabe:2015tja}, where the original CXY subtraction is used but a kinematical constraint is imposed which leads to an additional correction term $L_q$. This additional term extends the $\ptt$ range where the multiplicity is positive by about 0.5~GeV, which is similar to what was obtained in Ref.~\cite{Watanabe:2015tja} in the same kinematics.

This is the main result of our paper:
we have shown that it is possible to choose a value of $x_g/\xf$ which is still in its ``natural'' range and extends significantly the region where the NLO cross section has a reasonable physical behavior. In the case we have studied here the dependence on the exact choice for the renormalization scale is stronger than one would like, but we believe that this is due to two aspects of the calculation that can be improved in the future.
First, we tried here to implement ordering in an effective way by relying only on external scales to fix $\xif$. For more accurate results, one should instead impose the ordering~(\ref{eq:kminusorder}) inside the transverse momentum integrals in $\jcal$ and $\jcal_v$. This could lead to sizeable differences with the results we have shown since, as can be seen from Fig.~\ref{fig:dN_xifcontinuous}, our results are still very sensitive to the value of $\xif$. Second, we have used the GBW model to parametrize the dipole cross section. While this has practical advantages like enabling us to perform some integrals analytically, in this model the leading and next-to-leading order contributions have very different behaviors at large transverse momentum: the leading order term behaves like a gaussian, while the next-to-leading order corrections behave like a power law. Therefore, at large transverse momentum, the behavior of the multiplicity is governed entirely by the NLO corrections, which is quite unnatural. On the other hand, using a more physical dipole cross section, such as one obtained by solving the Balitsky-Kovchegov~\cite{Balitsky:1995ub,Kovchegov:1999yj} equation, should make the leading order contribution behave more like a power law. This means that the end result would be less sensitive to the exact choice made for the parametrization of $\xif$. Taking a BK-evolved dipole cross section instead of the GBW one would also make the cross section formally independent of the factorization scale $\xf$ at this order in $\as$, which should significantly reduce the factorization scale dependence.

\section{Discussion and outlook}
\label{sec:disc}

To summarize, we have in this paper proposed to modify the subtraction procedure of the rapidity divergence in the calculation of single inclusive hadron production in the hybrid formalism~\cite{Chirilli:2011km,Chirilli:2012jd}. It has been hoped that  imposing ordering in the light cone energy of the probe could solve the issue of large negative NLO corrections to the cross section. We have here argued that the most natural way to implement this ordering is to impose it on the kinematics of the part that is subtracted from the cross section and absorbed into the target. To explicitly demonstrate the effect of our proposal, we have performed the calculations  at finite $\nc$ for the GBW model. The part corresponding to the rapidity divergence is proportional to $\nc$, whereas  the parts associated with DGLAP evolution have a  color factor $\cf$. Compared to the CXY formulation, our suggestion only modifies the $\nc$-terms and leaves the $\cf$-terms unchanged. Indeed, we have shown that such a modification of the subtraction scale can lead to a more stable perturbative expansion for this cross section at high $\ptt$. By a suitable choice of the factorization scale $\xf$, 
the stability of the perturbative expansion can be extended to arbitrarily high $\ptt$. This should be contrasted with the recent calculation of Ref.~\cite{Watanabe:2015tja}, where a similar correction is obtained as an additional  correction which in some cases still leaves the cross section negative at high enough $\ptt$. The main difference between our approach here and that of  
Ref.~\cite{Watanabe:2015tja} is that there the kinematical modification is treated as a single correction term whereas here it is resummed to all orders by shifting the evolution variable in the BK equation.

In some works (e.g. \cite{Stasto:2014sea,Watanabe:2015tja}), the kinematical constraint  is explained as a requirement of keeping the momentum fraction in the target $x_a$ as less than $1$.  We would rather prefer to characterize this requirement  by saying that the $x_a$ of the corrections that are resummed into the BK evolution should be larger than the $x_g$ determined by the leading order kinematics. The whole formalism here is based on the eikonal approximation, which is valid for only small enough $x_a$. If the actual result for the cross section were to really depend on the dipole cross section at very large target $x_a$, the whole formalism would be in grave trouble. Fortunately this is not the case, since the contribution from large $x_a\sim 1$ (corresponding to $(1-\xi)\lesssim x_g$) is subtracted from the cross section and absorbed into the renormalization group evolution of the target: in fact it is only a tiny part of the subtraction. In spite of this different terminology, the actual equations in Refs.~\cite{Stasto:2014sea,Watanabe:2015tja} lead to the same parametric dependence that we have here: the longitudinal factorization scale has to be modified by the presence of a large transverse momentum logarithm $\sim \ln \ktt^2/\qs^2$.

In order to perform the whole calculation consistently at NLO accuracy, one needs to also solve the NLO version of the BK equation. To do this, the actual BK equation to be solved must be consistent with the subtraction procedure in the cross section. Different subtraction schemes correspond to different versions of the BK equation. Since the rapidity dependence of the dipole cross section is proportional to  $\as$,  a modification of the subtraction scale in \eq\nr{eq:dlzsub} is a higher order $\as^2$ effect, and the difference appears at the NLO level for the BK equation. In fact, it was recently shown in \cite{Iancu:2015vea} that a problematic double logarithmic correction to the NLO BK equation~\cite{Balitsky:2008zza} can indeed be resummed by redefining the evolution variable in a similar way as we have done in \eq\nr{eq:dlzsub}.  When written as a differential equation in $Y = \ln 1/\xf$, the modified BK equation is nonlocal in rapidity, which is inconvenient for a numerical solution. The authors of~\cite{Iancu:2015vea} proposed a clever way of rewriting this in a local form, which has successfully been solved numerically~\cite{Iancu:2015joa,Lappi:2016fmu}. 

Compared to the work in \cite{Iancu:2015vea,Iancu:2015joa,Lappi:2016fmu}, our proposal here has been a very simplistic one. By explicitly taking $\qs$ as the scale of the logarithm, we have been able to write a subtraction procedure in such a way that we can reuse most of the CXY formulae with only slight modifications. To be consistent \cite{Iancu:2015vea,Iancu:2015joa,Lappi:2016fmu}, the momentum scale in \eq\nr{eq:xifmin} should actually be one that is integrated over in  $\jcal$ and $\jcal_v$. This would require a new computation of the regularized ``hard factors'' remaining after the subtraction. For a calculation that is consistent with recent NLO BK solutions it would be useful  to carry out this subtraction for the resummation proposed in \cite{Iancu:2015vea}. This is, however, left for future work.

\section*{Acknowledgements} 
We thank E. Iancu, Z. Kang, B.-W Xiao and D. Zaslavsky for discussions. This work has been supported by the Academy of Finland, projects 
267321 and 273464.

\providecommand{\href}[2]{#2}\begingroup\raggedright\endgroup

\end{document}